\documentclass[reprint, prl, superscriptaddress]{revtex4-2} 
\usepackage[pdftex]{graphicx}
\usepackage{amsmath}
\usepackage{bm}

\begin{document}

\title{Magnetism of Kitaev spin-liquid candidate material RuBr$_3$}

\affiliation{Department of Physics, Graduate School of Science, Tohoku University, 6-3 Aramaki-Aoba, Aoba-ku, Sendai, Miyagi 980-8578, Japan}

\author{Yoshinori Imai}
\email{imai@tohoku.ac.jp}
\affiliation{Department of Physics, Graduate School of Science, Tohoku University, 6-3 Aramaki-Aoba, Aoba-ku, Sendai, Miyagi 980-8578, Japan}

\author{Kazuhiro Nawa}
\affiliation{Institute of Multidisciplinary Research for Advanced Materials, Tohoku University, Katahira, Sendai, 980-857, Japan}

\author{Yasuhiro Shimizu}
\affiliation{Department of Physics, Nagoya University, Furo-cho, Chikusa-ku, Nagoya 464-8602, Japan}

\author{Wakana Yamada}
\author{Hideyuki Fujihara}
\author{Takuya Aoyama}
\affiliation{Department of Physics, Graduate School of Science, Tohoku University, 6-3 Aramaki-Aoba, Aoba-ku, Sendai, Miyagi 980-8578, Japan}

\author{Ryotaro Takahashi}
\author{Daisuke Okuyama}
\affiliation{Institute of Multidisciplinary Research for Advanced Materials, Tohoku University, Katahira, Sendai, 980-857, Japan}

\author{Takamasa Ohashi}	
\affiliation{Department of Physics, Nagoya University, Furo-cho, Chikusa-ku, Nagoya 464-8602, Japan}


\author{Masato Hagihala}
\author{Shuki Torii}
\affiliation{Institute of Materials Structure Science, High Energy Accelerator Research Organization (KEK), 203-1 Tokai, Ibaraki, 319-1106, Japan}





\author{Daisuke Morikawa}
\author{Masami Terauchi}
\affiliation{Institute of Multidisciplinary Research for Advanced Materials, Tohoku University, Katahira, Sendai, 980-857, Japan}

\author{Takayuki Kawamata}
\author{Masatsune Kato}
\affiliation{Department of Applied Physics, Graduate School of Engineering, Tohoku University, Sendai 980-8579, Japan}

\author{Hirotada Gotou}
\affiliation{Institute for Solid State Physics, University of Tokyo, Kashiwa, Chiba 277-8581, Japan}

\author{Masayuki Itoh}
\affiliation{Department of Physics, Nagoya University, Furo-cho, Chikusa-ku, Nagoya 464-8602, Japan}

\author{Taku J. Sato}
\affiliation{Institute of Multidisciplinary Research for Advanced Materials, Tohoku University, Katahira, Sendai, 980-857, Japan}

\author{Kenya Ohgushi}
\affiliation{Department of Physics, Graduate School of Science, Tohoku University, 6-3 Aramaki-Aoba, Aoba-ku, Sendai, Miyagi 980-8578, Japan}

\newcommand{\rbb}{RuBr$_{3}$(chain)}
\newcommand{\rb}{RuBr$_{3}$}
\newcommand{\rbd}{RuBr$_{3}$ with the honeycomb structure}
\newcommand{\rbbd}{RuBr$_{3}$ with the chain structure}
\newcommand{\rc}{$\alpha$-RuCl$_{3}$}
\newcommand{\rcb}{$\beta$-RuCl$_{3}$}
\newcommand{\tc}{$T_\mathrm{c}$}
\newcommand{\tn}{$T_\mathrm{N}$}
\newcommand{\tp}{$T_\mathrm{peak}$}
\newcommand{\ts}{$T^*$}
\newcommand{\eg}{$E_\mathrm{g}$}

\begin{abstract}
The ruthenium halide \rc~is a promising candidate for a Kitaev spin liquid. 
However, the microscopic model describing \rc~is still debated partly because of 
a lack of analogue materials for \rc, which prevents tracking of electronic properties as functions of controlled interaction parameters.
Here, we report a successful synthesis of \rb.
The material \rb~possesses BiI$_3$-type structure (space group: $R\overline{3}$) where Ru$^{3+}$ form an ideal honeycomb lattice. 
Although \rb~ has a negative Weiss temperature, it 
undergoes a zigzag antiferromagnetic transition at $T_\mathrm{N}=34$ K, as does \rc.
Our analyses indicate that the Kitaev and non-Kitaev interactions can be modified in ruthenium trihalides by changing the ligand sites, which provides a new platform for exploring Kitaev spin liquids.

\end{abstract}

\maketitle

\begin{figure*}[th]
\begin{center}
\begin{minipage}{16.5cm}
\includegraphics[width=1\linewidth]{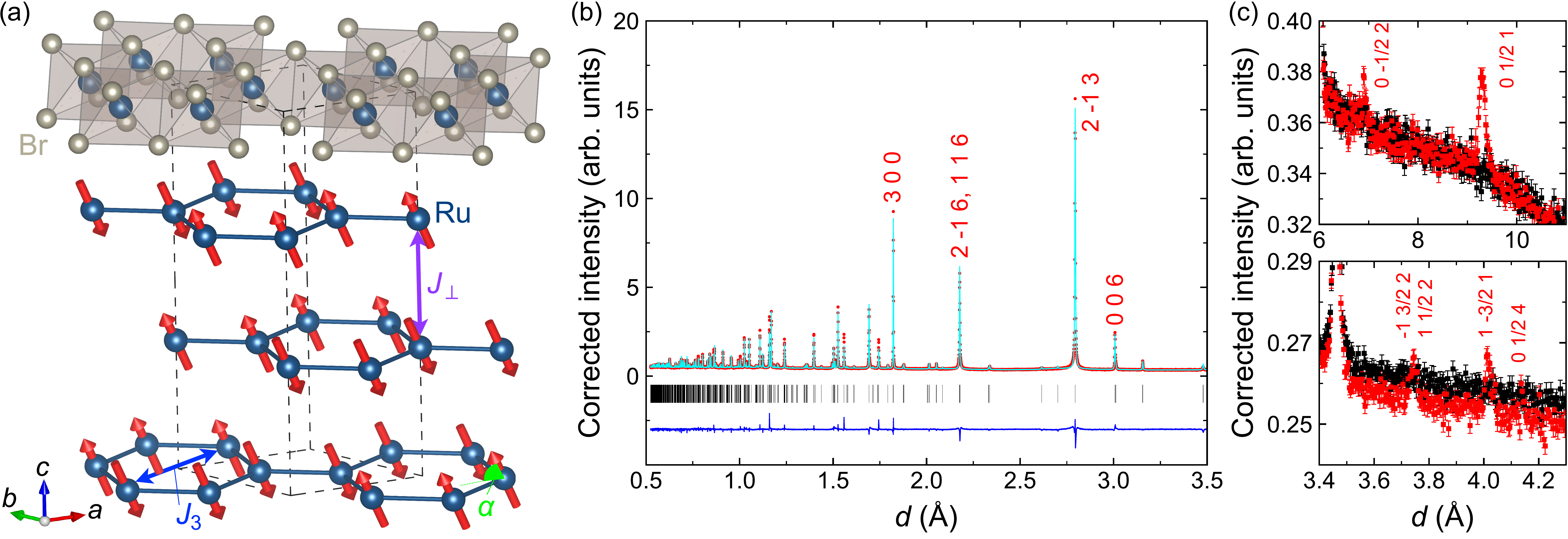}
\caption{
Crystal and magnetic structures of RuBr$_3$ with the BiI$_3$ structure. 
(a) Schematic of crystal and magnetic structures refined from the neutron diffraction pattern.
This figure is drawn using VESTA \cite{vesta}.
$\alpha$ is the angle of the magnetic moment from the honeycomb plane, and $J_3$ and $J_\perp$ are the third-nearest-neighbor and the interlayer Heisenberg interactions, respectively. 
(b) Neutron diffraction pattern collected at 300~K by the backscattering (BS) bank. The red dots, cyan curve, and blue curve represent the observed intensities, calculated intensities, and difference between them. The difference is shifted for clarity.
The vertical black bars indicate the peak position expected from the space group $R\overline{3}$.
(c) Neutron diffraction patterns focused on the magnetic reflections. Intensities at 3 and 100~K are represented by the red and black dots, respectively. The upper and lower panels are data collected by the quarter-angle (QA) and low-angle (LA) detector banks, respectively.
}
\label{fig:xrd}
\end{minipage}
\end{center}
\end{figure*}

A theoretical breakthrough \cite{kitaev} has brought compounds with honeycomb lattices into the spotlight as candidates for quantum spin liquids.
Kitaev proposed a model where $S=1/2$ spins placed on a honeycomb lattice are coupled with their three nearest-neighbor spins with bond-dependent ferromagnetic Ising interactions \cite{kitaev}.
This model is exactly solvable, and the ground state is a quantum spin liquid.
%
The material \rc~is the most promising candidate for a Kitaev spin liquid and has been actively studied in recent years \cite{PRL.102.017205,JPCM.29.493002,JPSJ.89.012002}.
In \rc, Ru$^{3+}$ form a honeycomb lattice in two-dimensional layers through an edge-sharing network of RuCl$_6$ octahedra.
The detailed crystal structure of \rc~is debated \cite{RuCl3_0,PRB.92.235119, PRB.93.134423, NatMat.15.733}. 
Recent studies on high-quality single crystals report that it has an AlCl$_3$-type crystal structure at room temperature with a slightly distorted honeycomb lattice
(space group: $C2/m$), and that below a structural transition temperature of 150 K, it has a BiI$_3$-type structure with an ideal honeycomb lattice (space group: $R\overline{3}$, Fig.~\ref{fig:xrd}(a)) \cite{PRB.91.094422,arxiv160905690,PRB.101.020414}.
Instead of the expected spin-liquid state owing to the frustrated nature of the Kitaev interaction $K$,
\rc~shows a zigzag antiferromagnetic order \cite{PRB.92.235119}.
The antiferromagnetic transition temperature, $T_\mathrm{N}=7-14$ K, is very sensitive to the details of the crystal structure \cite{PRB.91.144420,PRB.93.134423}. 
The long-range antiferromagnetic order originates from the sizable contributions of non-Kitaev interactions such as the Heisenberg interaction $J$ and off-diagonal interaction $\Gamma$.
In fact, the extended model that takes such contributions into account $($the $J$-$K$-$\Gamma$ model$)$ is useful for describing magnetic properties of \rc.
The Hamiltonian is written as
\begin{equation}
\mathcal{H}=\sum_{\left < i, j \right > \in NN}{\left ( J\bm{S}_i \cdot \bm{S}_j + K S_i^\gamma S_j^\gamma + \Gamma \left ( S_i^{\alpha} S_j^{\beta} + S_i^{\beta} S_j^{\alpha} \right ) \right )},
\label{eq:jkg}
\end{equation}
where $S_{i}^{\gamma}$ denotes the $\gamma$ component of the spin-1/2 operator of site $i$ ($\gamma$ being the direction perpendicular to the edge-sharing plane) \cite{PRL.112.077204,PRB.92.024413,PRB.96.064430}.
By assuming appropriate values of the parameters $J$, $K$, and $\Gamma$, 
this Hamiltonian (eq. \ref{eq:jkg}) can well explain various magnetic behaviors of \rc, including the zigzag antiferromagnetic order, the field-induced Kitaev spin-liquid state, and magnetic excitations in Raman and inelastic neutron spectra \cite{PRL.112.077204,PRL.114.147201,NatPhys.12.912,Nat.559.227}.
Earlier works have revealed a dominant ferromagnetic Kitaev interaction of $K = -3$ -- $-25$ meV and sizable contributions from the other terms, $\left | J/K \right | \sim 0.05$ -- $0.25$ and $\left | \Gamma/K \right | \sim 0.21$ -- $0.99$ \cite{NQM.5.2}.
Hence, \rc~is still far from the ideal Kitaev limit, $\left | K \right | \gg \left | J \right |,~\left | \Gamma \right |$.
However, no methods to tune the parameters $J$, $K$, and $\Gamma$ have been established experimentally, and no state closer to the Kitaev limit has been achieved yet. This is partly because there are no analogue materials for \rc.

Here, we focus on two polymorphs of RuCl$_3$: an $\alpha$ one with a honeycomb structure and a $\beta$ one with a one-dimensional chain structure~(space group: $P6_3/mcm$) \cite{ZAAC.630.2199}.
While both polymorphs exist for RuCl$_3$, only the chain structure is known for RuBr$_3$ \cite{ZAAC.630.2199,JLCM.11.288}.
Because \rc~is more densely packed than \rcb, one expects that \rbd~can be obtained by keeping \rbbd~under high pressure and high temperature.
Actually, recent first-principles calculations predict the stability of \rbd, which can be foiled into a monolayer limit \cite{JMMM.476.111}.
Replacing Cl with Br is expected to have two effects: lattice expansion owing to the larger ionic radius of Br$^-$ than that of Cl$^-$, and increased covalency of Ru-Cl/Br bonds owing to the up-floating of Br: $4p$ bands compared with Cl: $3p$ bands.
As will be discussed later, both effects bring the system closer to the $\left | K \right | \gg \left | J \right |,~ \left | \Gamma \right |$ regime.


The X-ray diffraction pattern of RuBr$_3$ synthesized at high pressure is totally different from that of \rbbd~(see Supplementary Fig. S1).
To clarify the crystal structure, we measured the high-resolution neutron powder diffraction at various temperatures; the pattern collected at 300~K is shown in Fig.~\ref{fig:xrd}(b).
Among several crystal structures with different stacking sequences of the transition-metal trihalides \cite{AX3_0, ZKCM.176.233, JAC.246.70, Cryst.7.121}, the BiI$_3$ structure with the space group $R\overline{3}$ (Fig.~\ref{fig:xrd}(a)) best reproduces the experimental data.
We cannot find any signatures of a structural transition such as peak splitting down to 3~K.
The crystallographic data obtained through the Rietveld analysis are summarized in the supplementary information.
In the refined structure, the Ru atoms in the unit cell are related to each other through the threefold rotation axis along the $c$-direction, and the network formed by the nearest-neighbor Ru--Ru bonds is a regular honeycomb lattice.
This indicates that the newly developed \rb~provides an ideal platform for exploring Kitaev spin liquids.
Henceforth, we discuss the electronic properties of \rb~with the $R\overline{3}$ structure.

\begin{figure}[th]
\begin{center}
\includegraphics[width=0.85\linewidth]{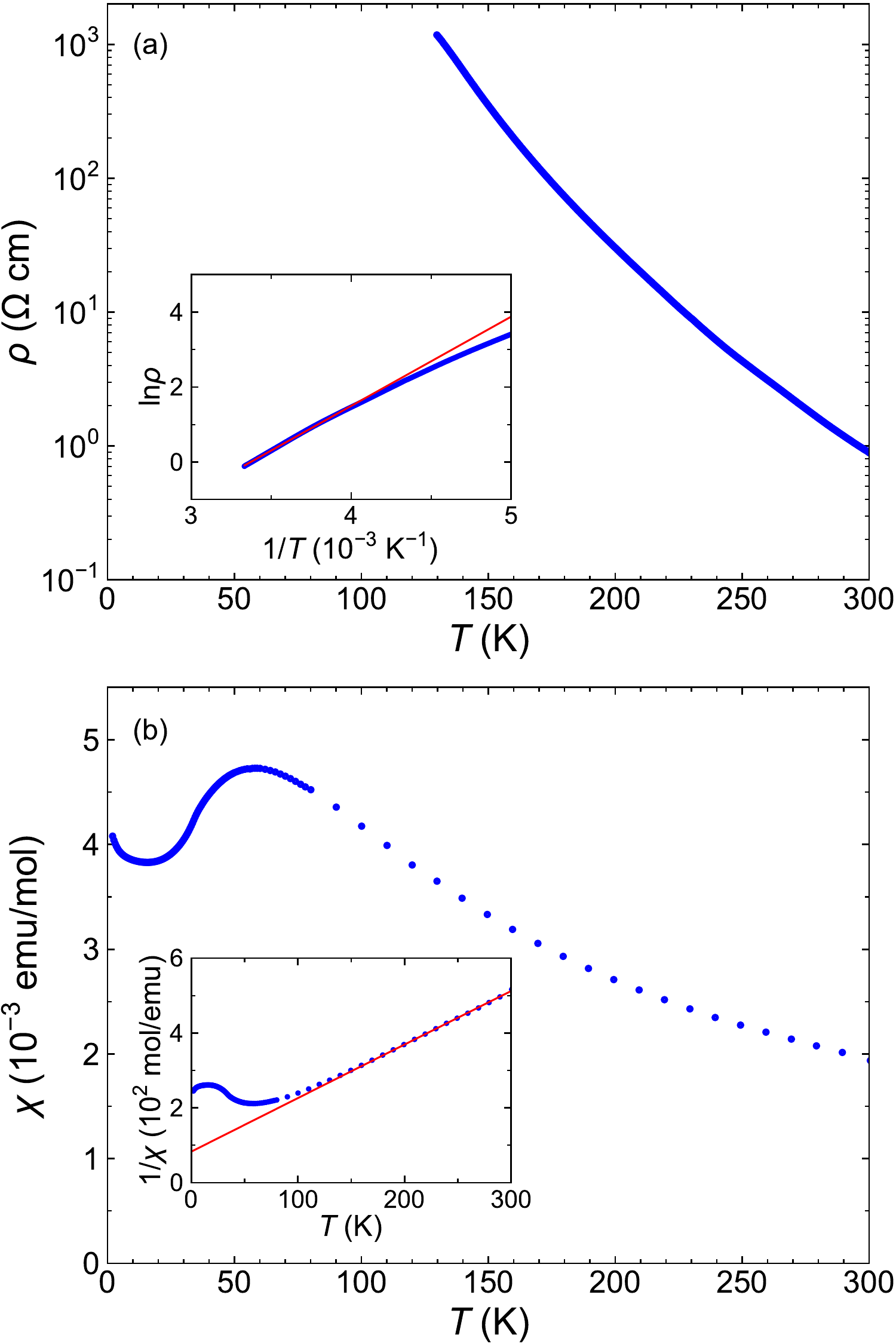}
\caption{
Basic electronic properties of RuBr$_3$ with the BiI$_3$ structure.
(a) Temperature $(T)$ dependence of the resistivity $(\rho)$. 
The inset shows the Arrhenius plot, and the fitting result is shown by the red line.
(b) Temperature dependence of the magnetic susceptibility ($\chi$) under a magnetic field of $\mu_0 H = 1$ T.
The inset shows the inverse of $\chi$.
The data at $200-300$ K are fitted with the Curie-Weiss law, and the fitting results are shown by the red line.
}
\label{fig:rt}
\end{center}
\vspace{-5mm}
\end{figure}
%
%
\begin{figure}[th]
\begin{center}
\includegraphics[width=0.85\linewidth]{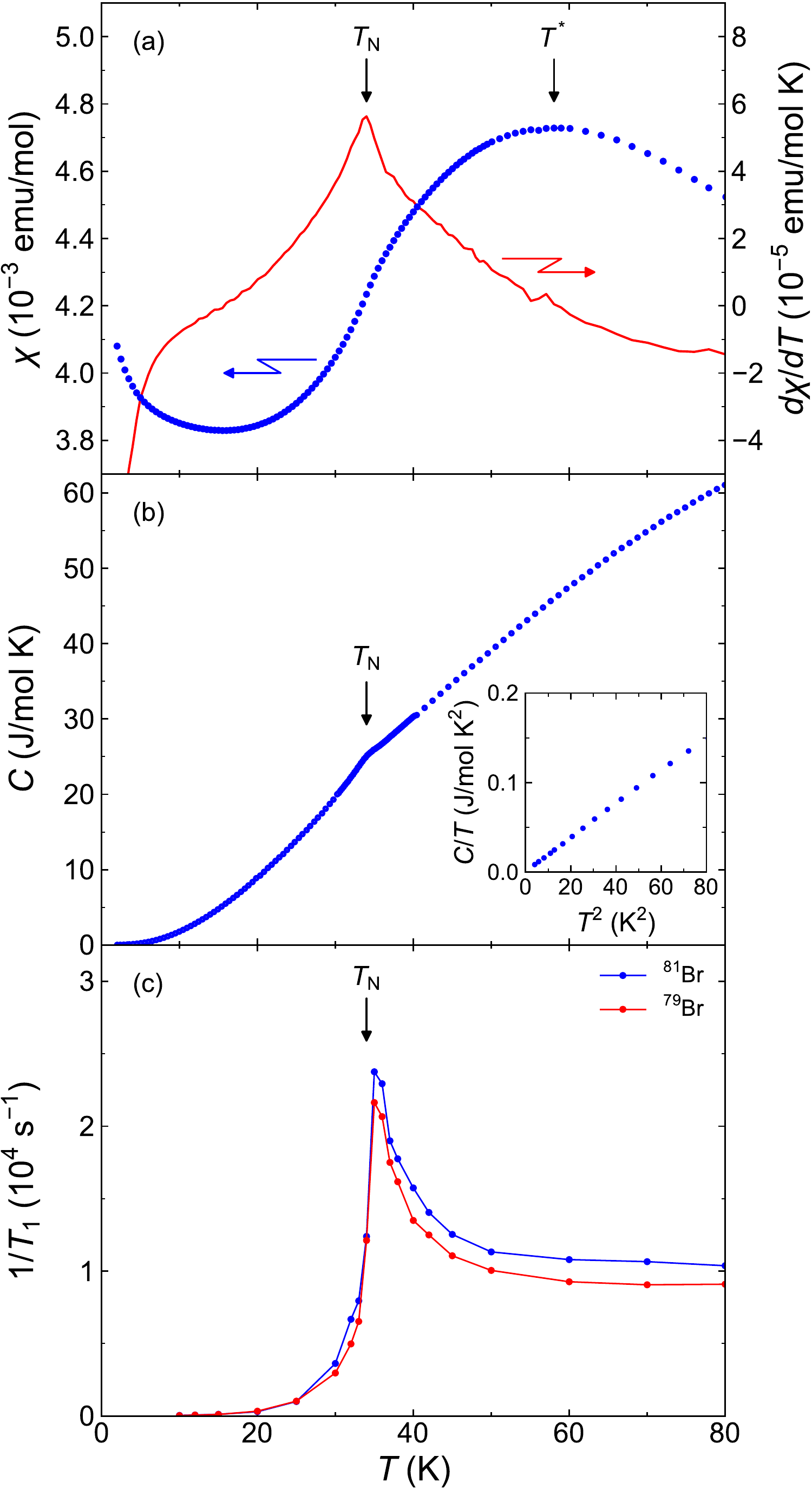}
\caption{
Electronic properties near the antiferromagnetic transition for RuBr$_3$ with the BiI$_3$ structure.
(a-c) Temperature $(T)$ dependences of (a) the magnetic susceptibility ($\chi$) and temperature differential of $\chi$ $(d\chi/dT)$ under a magnetic field of $\mu_0 H = 1$ T, (b) the specific heat $(C)$ under a zero magnetic field, and (c) the $^{79}$Br and $^{81}$Br nuclear spin-lattice relaxation rate $1/T_1$. 
In the inset of (b), $C/T$ is plotted as a function of $T^2$.
The solid curves in (c) are guides for the eye.
}
\label{fig:lt}
\end{center}
\vspace{-5mm}
\end{figure}

Figure \ref{fig:rt}(a) shows the temperature dependence of the resistivity $\rho$ for \rb.
The data show a thermally activated temperature dependence, which is consistent with a strongly spin-orbital-coupled Mott insulator.
The activation energy is $E_g \sim 0.21$ eV.
Figure \ref{fig:rt}(b) shows the temperature dependence of the magnetic susceptibility $\chi$ under a magnetic field of $\mu_0 H = 1$ T.
The data at $200-300$ K well follow Curie-Weiss behavior, $\chi = C_\mathrm{CW}/(T-\theta_\mathrm{CW})$, with a Curie constant of $C_\mathrm{CW}=0.699$ emu/mol and a Weiss temperature of $\theta_\mathrm{CW}=-58$ K.
The effective moment per ruthenium ion, $\mu_{eff}$, is calculated to be $2.36~\mu_\mathrm{B}$ from the relation $C_\mathrm{CW}=N_\mathrm{A} \mu_{eff}^2 /3 k_\mathrm{B}$, where $N_\mathrm{A}$ and $k_\mathrm{B}$ are the Avogadro constant and the Boltzmann constant, respectively.
The negative Weiss temperature indicates the predominance of the antiferromagnetic correlation in this system.
Turning to the low-temperature regime, $\chi$ shows a broad peak around $T^* = 60$ K, which is most likely related to the development of antiferromagnetic correlations.
When the system is cooled further, $\chi$ shows a kink characterized by a sharp peak in $d\chi/dT$ at $T_\mathrm{N}=34$ K, as shown in Fig. \ref{fig:lt}(a).
This kink anomaly corresponds to the formation of long-range antiferromagnetic order.
This interpretation is supported by the result for the specific heat $C$, which shows a peak just around 34 K as shown in Fig. \ref{fig:lt}(b).
The low-temperature part of $C$ obeys a $T^3$ law $(C=\beta T^3)$ as shown in the inset of Fig. \ref{fig:lt}(b), which can be explained by contributions from Debye phonons and antiferromagnetic spin waves in three dimensions. If we suppose that spin-wave contributions are negligible, we obtain a Debye temperature of $\theta_\mathrm{D} = 159$ K from the formula $\beta=12\pi^4NR/5\theta_\mathrm{D}^3$, where $N$ is the number of atoms per formula unit and $R$ is the gas constant.

\begin{figure}[b]
\begin{center}
\includegraphics[width=0.85\linewidth]{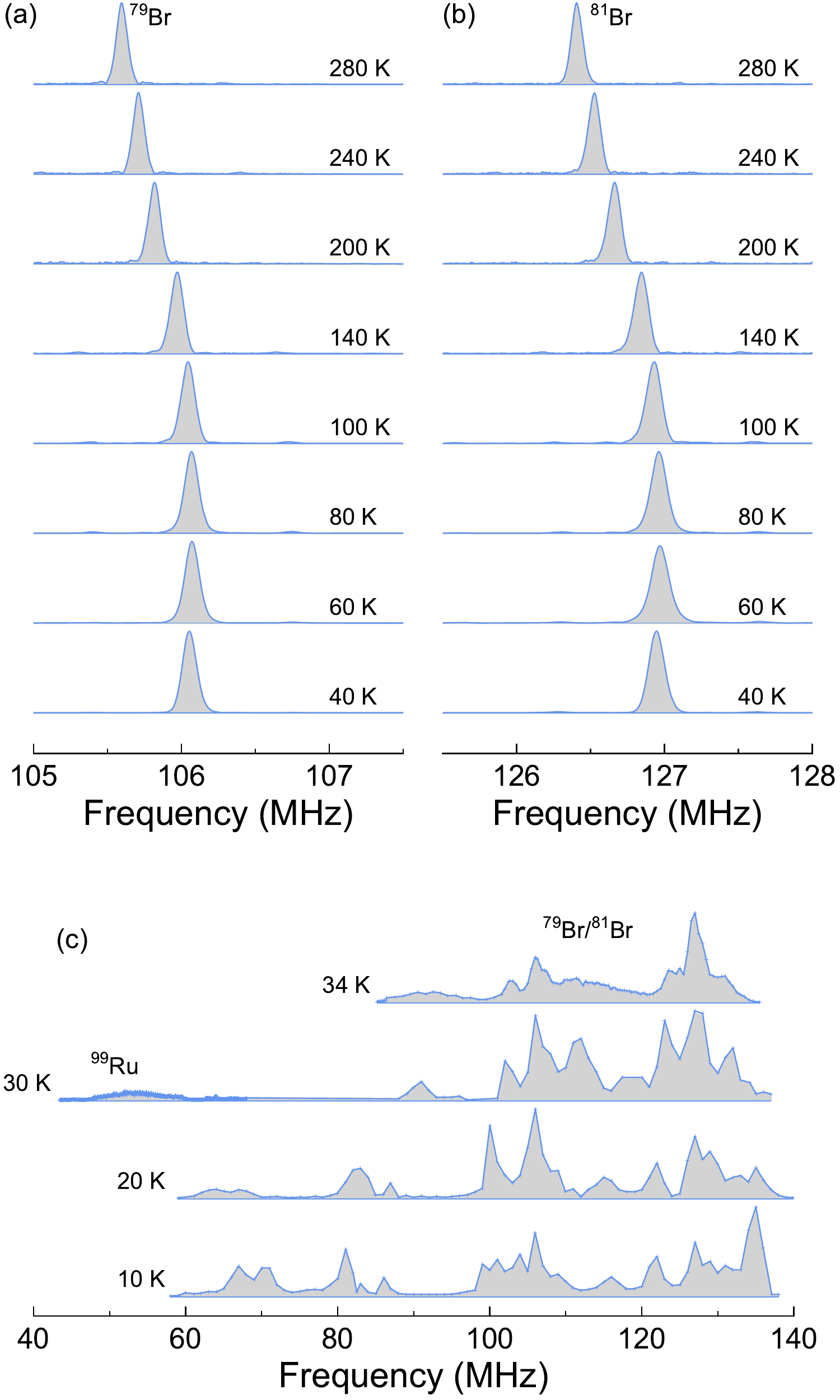}
\caption{
Nuclear quadrupole resonance (NQR) spectra for RuBr$_3$ with the BiI$_3$ structure.
(a, b) NQR spectra of (a) $^{79}$Br and (b) $^{81}$Br above the antiferromagnetic transition temperature.
(c) Zero-field Br and Ru nuclear magnetic resonance (NMR) spectra in the antiferromagnetically ordered state.
A spectrum below 75 MHz comes from a $^{99}$Ru NMR spectrum.
}
\label{fig:nqr}
\end{center}
\vspace{-5mm}
\end{figure}

To investigate the spin dynamics, we performed nuclear quadrupole resonance (NQR) spectroscopy.
Figures \ref{fig:nqr}(a) and (b) show NQR spectra of $^{81}$Br and $^{79}$Br (nuclear spin $I = 3/2$) for \rb~at $40-280$ K. 
The resonance frequency scales well with the electric quadrupole moment ($^{81}Q = 0.276b$, $^{79}Q = 0.330b$, $b= 10^{-28}$ m$^2$). 
A single NQR spectrum for each nucleus is consistent with an $R\overline{3}$ structure that contains one Br site. 
The temperature dependence of the NQR frequency reflects an enhancement of the electric field gradient via thermal lattice contraction. 
Upon further cooling, the NQR spectrum splits below 34 K, clearly showing the emergence of a spontaneous local field due to the antiferromagnetic order. 
Figure \ref{fig:lt}(c) shows the temperature dependence of the $^{79}$Br and $^{81}$Br nuclear spin-lattice relaxation rate $1/T_1$. 
The values for both isotopes scale well with the square of the gyromagnetic ratio, as expected for predominant spin fluctuations. 
An increase in $1/T_1$ from about 60 K shows the growth of antiferromagnetic correlation; this is in harmony with the broad peak structure at $T^*$ in $\chi$. 
The spin fluctuations divergently increase toward \tn, below which magnetic excitation is gapped, coinciding with the evolution of the order parameter as observed in the zero-field nuclear magnetic resonance (NMR) spectrum (Fig. \ref{fig:nqr}(c)).

To reveal the magnetic structure, we performed neutron diffraction measurements at low temperature, and the collected patterns are shown in Fig.~\ref{fig:xrd}(c).
At least five peaks are found at 3~K, while there are none at 100~K.
These peaks can be indexed with the magnetic modulation vector $\mathbf{k}=(0,1/2,1)$.
Initial candidates for determining the magnetic structure were obtained using magnetic representation theory \cite{JMMM.12.239}.
The magnetic representations for the Ru moments were decomposed into two one-dimensional representations (IR1 and IR2) of the $\mathbf{k}$-group with the magnetic modulation vector $\mathbf{k} = (0, 1/2, 1)$ (see the supplementary information for the details of the IRs). 
The powder neutron diffraction pattern was well fit with the IR2 representation, which represents a zigzag antiferromagnetic structure (Fig.~\ref{fig:xrd}(a)).
A zigzag order at low temperature in \rb~agrees with the theoretical prediction, which takes electron correlation into account \cite{JMMM.476.111}.
The magnetic moment is tilted, and its angle from the $ab$-plane is $\alpha=64(12)^\circ$ (Fig. \ref{fig:xrd}(a)).
The magnitude of the magnetic moment is estimated to be $m=0.74(12)~\mu_\mathrm{B}$.

We here compare the experimental results for two honeycomb-based materials, \rb~and \rc.
The material \rb~has the $R\overline{3}$ structure in the whole temperature range below room temperature, contrasting with the structural transition at 150 K from the $C2/m$ to the $R\overline{3}$ structure in \rc~\cite{PRB.91.094422,arxiv160905690,PRB.101.020414}.
The metal-metal bond distance is $d_\mathrm{Ru-Ru} = 3.6438(2)~\mathrm{\AA}$ in \rb, which is much longer than the reported values $d_\mathrm{Ru-Ru}=3.43$ and $3.46~\mathrm{\AA}$ in \rc~\cite{PRB.92.235119,PRB.93.134423,yi19}.
The interlayer distance between the honeycomb planes is $d_\perp = 6.014(1)~\mathrm{\AA}$ in \rb, which is also longer than the reported $d_\perp=5.72~\mathrm{\AA}$ in \rc~\cite{PRB.92.235119,PRB.93.134423,yi19}.
These are the consequences of the larger ionic radius of Br$^-$ than that of Cl$^-$.
The Ru-Cl/Br-Ru bond angle, which is an indicator of trigonal distortion, is closer to the ideal value of $90^\circ$ for \rb~$(\phi = 93.14(8)^\circ)$ than for \rc~$(\phi \sim 94^\circ)$ \cite{PRB.92.235119,PRB.93.134423}.
Hence, the $J_{eff}=1/2$ state is still a good starting point in \rb, which is also evidenced by the comparable effective magnetic moment: $\mu_{eff}=2.36~\mu_\mathrm{B}$ for \rb~and $\mu_{eff}=2.24~\mu_\mathrm{B}$ for \rc.
The crucial difference between the two materials is the dominant magnetic interaction.
In \rb, there are fairly strong antiferromagnetic interactions characterized by a negative Weiss temperature of $\theta_\mathrm{CW}=-58$ K; in \rc, there are ferromagnetic interactions characterized by a positive Weiss temperature of $\theta_\mathrm{CW}=19$ K.
Nevertheless, both compounds have zigzag magnetic order at low temperature.
The magnetic order in \rb~is more robust than in \rc.
This can be seen from the differences in \tn~$(T_\mathrm{N}=34$ K for \rb~and $T_\mathrm{N}=7-14$ K for \rc$)$ and the ordered magnetic moment $(m=0.74(12)\mu_\mathrm{B}$ for \rb~and $m=0.4$--$0.7~\mu_\mathrm{B}$ for \rc~\cite{PRB.92.235119, PRB.93.134423, arxiv160905690, NatMat.15.733}$)$.
The tilt angle of the magnetic moment from the honeycomb plane is $\alpha=64(12)^\circ$ in \rb, which is larger than $\alpha=32$--$35^\circ$ in \rc~\cite{PRB.93.134423,NatPhys.16.837}. 
Concerning magnetic fluctuations above \tn, both materials show an enhancement in $1/T_1$ toward \tn~below nearly the same temperature, $\sim 60$ K \cite{PRB.101.020414}.
However, the broad peak structure in the $\chi$ curve in this temperature regime is only seen in \rb. 


We now discuss the microscopic origin of the similarity/difference in magnetic properties of \rb~and \rc~on the basis of the $J-K-\Gamma$ model (eq. \ref{eq:jkg}).
The $J-K-\Gamma$ model contains the zigzag magnetic order as the ground state in a wide parameter space.
Let us first discuss how the replacement of Cl with Br influences $J$, $K$, and $\Gamma$.
The leading terms of $J$, $K$, and $\Gamma$ deduced from perturbation theory can be written as follows \cite{JPSJ.83.093701,PRB.90.045102,PRL.112.077204,PRB.93.214431,JPCM.32.404001}:
\begin{equation}
J \simeq \frac{4}{9}\frac{t_{dd}^2}{U},~K \simeq -\frac{8}{3}\frac{t_{dpd}^2}{U}\frac{J_\mathrm{H}}{U},~\Gamma \simeq -\frac{16}{9}\frac{t_{dpd}t_{dd}}{U}\frac{J_\mathrm{H}}{U},
\label{eq:j}
\end{equation}
where $U$ and $J_\mathrm{H}$ are the Coulomb repulsion and the Hund coupling between $t_{2g}$ electrons at the same site, respectively.
These exchange interactions are sensitive to the direct hopping between nearest-neighbor $t_{2g}$ orbitals, $t_{dd}$, and the indirect hopping between nearest-neighbor $t_{2g}$ orbitals via ligand $p$ orbitals, $t_{dpd}$.
Replacing Cl with Br brings about drastic change especially in $t_{dd}$ and $t_{dpd}$.
Because
$d_\mathrm{Ru-Ru}$ for \rb~is much longer than for \rc, replacing Cl with Br results in a significant decrease in $t_{dd}$.
The $t_{dpd}$ values would be enhanced by replacing Cl with Br, because Br $4p$ orbitals are more strongly hybridized with Ru $4d$ orbitals than Cl $3p$ orbitals.
As a result, \rb~is expected to have a higher $\left | K \right |$ and lower $\left | J/K \right |$ and $\left | \Gamma/K \right |$ than \rc.
Therefore, replacing Cl with Br drives the system closer to the Kitaev spin-liquid limit.
This is clearly evidenced by the tilt angle of the magnetic moment from the honeycomb plane, $\alpha$.
Theoretical calculations based on the $J-K-\Gamma$ model indicate that $\alpha$ monotonically increases by approaching the Kitaev spin-liquid limit in the zigzag antiferromagnetic phase \cite{PRB.92.024413,PRB.93.155143,PRB.94.064435,PRB.99.064425}.
Our experimental observation of larger $\alpha$ in \rb~than in \rc~is persuasive evidence that \rb~is closer to the Kitaev spin-liquid limit than \rc.

This then raises a new question of why the zigzag antiferromagnetic order in \rb~is more stable than in \rc, even though \rb~is closer to the Kitaev spin-liquid limit with much geometrical frustration.
One possible answer is that the energy scale of the $J-K-\Gamma$ model becomes much larger in \rb~than in \rc; 
however, this idea does not account for the difference in sign of the Weiss temperature between \rb~and \rc.
The other plausible reason for the stable magnetic order in \rb~is obtained by going beyond the $J-K-\Gamma$ model. 
The negative Weiss temperature in \rb~implies a significant contribution from the antiferromagnetic non-Kitaev interactions.
The possible candidates are the third-nearest-neighbor Heisenberg interaction $J_3(>0)$ and the interlayer coupling $J_\perp (>0)$, which are shown in Fig. \ref{fig:xrd}(a).
These exchange interactions are believed to play an important role in the formation of zigzag antiferromagnetic order \cite{PRB.93.214431,PRL.112.077204,NJP.16.013056,PRL.113.107201,PRB.101.174444}.
The hopping processes responsible for both $J_3$ and $J_\perp$ occur through Ru$4d-$Br$4p-$Br$4p-$Ru$4d$ paths \cite{JPCM.29.493002, PRB.101.174444},
which are pronounced owing to the enhanced hopping integrals between Ru$4d-$Br$4p$ and Br$4p-$Br$4p$ orbitals.
Let us estimate the energy scale of $J_3$ and $J_\perp$.
In the mean-field approximation, the Weiss temperature for a powder-averaged system is \cite{PRB.103.L220408,PRB.101.174444}
\begin{equation}
k_\mathrm{B}\theta_\mathrm{CW} = -\frac{3}{4} \left ( J + \frac{1}{3}K + J_3 + \frac{1}{3}J_\perp \right ).
\label{eq:Tcwth}
\end{equation}
If we suppose that $J$ is negligible, we can conclude from the negative Weiss temperature that $\left | 3 J_3 + J_\perp \right |$ is at least comparable to $K$.
Our argument based on angle $\alpha$ in the preceding paragraph is not altered considerably, because it is not sensitive to the third-nearest-neighbor Heisenberg interaction \cite{PRB.94.064435,PRB.99.064425}.
Validation of our approach requires experimental studies using \rb~single crystals and support from theoretical studies.

In conclusion, we successfully synthesized \rb~with a BiI$_3$-type structure (space group: $R\overline{3}$), where Ru$^{3+}$ form an ideal honeycomb lattice.
\rb~shows a zigzag antiferromagnetic transition at $T_\mathrm{N}=34$ K, which is significantly higher than $T_\mathrm{N}=7-14$ K in \rc.
Our results indicate that the Kitaev and non-Kitaev interactions can be modified in ruthenium trihalides by changing the ligand sites, and provide a new platform for exploring Kitaev spin liquids. \\


%
%
%
%
%

\noindent {\bf Methods}


The \rb~polycrystalline sample with the BiI$_3$ structure (space group: $R\overline{3}$) was synthesized using a cubic-anvil high-pressure apparatus.
The starting material, commercially available \rbbd~(space group: $P6_3/mcm$), was placed in a gold or platinum capsule and loaded into a pyrophyllite cube.
These were pressurized at $\sim4$ GPa and heated at $\sim400^\circ$C for 30 minutes.
The electrical resistivity was measured using the four-terminal method over the temperature range $2-300$ K.
Magnetic susceptibility measurements were performed using a superconducting quantum interference device (SQUID) magnetometer.
The specific heat was measured using the thermal-relaxation method at temperatures as low as 2 K under a magnetic field of $\mu_0 H=0-9$ T using a commercial apparatus (Physical Property Measurement System, Quantum Design). 


Powder neutron diffraction experiments were performed using the high-resolution time-of-flight (TOF) neutron powder diffractometer SuperHRPD installed at the beam line BL08 of J-PARC \cite{JPSJ.80SB.SB020}. 
A powder sample of $\sim 1.4$ g was loaded into a vanadium-nickel cylindrical sample can with a diameter of 6~mm.
The sample can was then put in a $^4$He closed-cycle refrigerator with the lowest attainable temperature of 3 K.
SuperHRPD has three sets of detector banks: backscattering (BS), quarter-angle (QA), and low-angle (LA).
The LA detector bank, which is in the low $Q$ range, has the advantage of detecting magnetic reflections,
while the BS detector bank, which achieves high resolution, is useful for investigating crystal structure precisely. 
The observed intensities were corrected for neutron absorption numerically.
Rietveld analyses were done using the Fullprof software suite \cite{PhysB.192.55}.

All three diffraction patterns were simultaneously used with the same weighting in the crystal-structure refinement,
while the diffraction patterns from the QA and LA detector banks with a $d$-range larger than $2.9~\mathrm{\AA}$ were used for the magnetic-structure refinement.
A pseudo-Voigt function convoluted with back-to-back exponentials \cite{JAC.15.581} was used to fit the peak profile.
In addition, a phenomenological model of the anisotropic peak broadening induced by the strain was applied to the peak width \cite{JAC.32.281},
because the peak width cannot be described as a smooth function of the $d$-spacing.
The Ru$^+$ form factor was used for the magnetic structure refinement because the Ru$^{3+}$ form factor was not available. 


Nuclear quadrupole resonance (NQR) of $^{79}$Br and $^{81}$Br was measured for $40-290$ K.
The NQR spectra were measured with the spin-echo method with the pulse interval $\tau = 15-20$ $\mu$s and $\pi/2$ duration $= 1.5-2$ $\mu$s. 
The NQR frequencies and intensities for $^{79}$Br and $^{81}$Br were consistent with their nuclear spins $(^{79,~81}I= 3/2)$, electric quadrupole moments $(^{79}Q = 0.33b$, $^{81}Q = 0.28b)$, and natural abundances $(^{79}W = 50\%,~^{81}W = 50\%)$. 
At zero field, the nuclear spin-lattice relaxation rate $1/T_1$ was measured with the saturation recovery and evaluated by fitting the recovery curve into a single exponential function.



\providecommand{\noopsort}[1]{}\providecommand{\singleletter}[1]{#1}

\   \\ 

\noindent {\bf Acknowledgements}

We would like to thank Yohei Yamaji, Takahiro Misawa, Hakuto Suzuki, Joji Nasu, and Yukitoshi Motome for fruitful discussions, and Takumi Hiraoka and Takeshi Yajima for their help in the X-ray diffraction measurements.
This work was carried out under the Visiting Researcher's Program of the Institute for Solid State Physics, the University of Tokyo.

This work was financially supported by JSPS KAKENHI under Grant Numbers JP18H01159, JP18K03531, JP19H01837, JP19H04685, JP19H05822, JP19H05823, JP19H05824, JP19K21837, and JP20H01850; by the Murata Science Foundation;
and by JST CREST under Grant Number JP19198318. 


\end{document}